%% file: paper.tex
\documentclass[12pt]{article}

\usepackage{sbc-template}
\usepackage{graphicx,url}
\usepackage[utf8]{inputenc}
\usepackage[english]{babel}  

\input{packages}

\input{macros}

\title{Securing IoT Apps with Fine-grained Control of Information Flows}

\author{Davino Mauro Junior\inst{1}, Kiev Gama\inst{1}, Atul Prakash\inst{2}}

\address{Centro de Informática (CIn) -- Universidade Federal de Pernambuco
  (UFPE)\\
  Recife, PE, Brazil
\nextinstitute
  Department of Electrical Engineering and Computer Science -- University of Michigan\\
  Ann Arbor, Michigan, U.S.
\email{\{dmtsj,kiev\}@cin.ufpe.br, aprakash@umich.edu\}}
}

\begin{document} 

\maketitle

\begin{abstract}
  Internet of Things is growing rapidly, with many connected devices
  now available to consumers. With this growth, the IoT apps that
  manage the devices from smartphones raise significant security
  concerns. Typically, these apps are secured via sensitive
  credentials such as email and password that need to be validated
  through specific servers, thus requiring permissions to access the
  Internet. Unfortunately, even when developers of these apps are
  well-intentioned, such apps can be non-trivial to secure so as to
  guarantee that user’s credentials do not leak to unauthorized
  servers on the Internet. For example, if the app relies on
  third-party libraries, as many do, those libraries can potentially
  capture and leak sensitive credentials. Bugs in the applications can
  also result in exploitable vulnerabilities that leak
  credentials. This paper presents our work in-progress on a prototype
  that enables developers to control how information flows within the
  app from sensitive UI data to specific servers. We extend FlowFence
  to enforce fine-grained information flow policies on sensitive UI
  data.
\end{abstract}

\input{intro}

\input{motivation}
\input{related}

\input{background-flowfence}
\input{approach}
\input{conclusion}
\input{acknowledgements}

\bibliographystyle{sbc}
\bibliography{paper}

\end{document}

%% file: packages.tex
\usepackage[english]{babel}
\usepackage{textcomp}
\usepackage{float}
\usepackage{listings}
\usepackage{caption}
\usepackage{graphicx}
\usepackage{subcaption}
\usepackage{colortbl}
\usepackage{amsmath} 
\usepackage{mathpartir}
\usepackage{fancyvrb}\fvset{fontsize=\small}
\usepackage{xspace}
\usepackage[table]{xcolor}
\usepackage{balance}
\usepackage{wrapfig}
\usepackage{multirow}
\usepackage{pifont}
\usepackage{amssymb}
\usepackage{soul}
\usepackage{microtype}
\usepackage[T1]{fontenc}
\usepackage{dsfont}
\usepackage{relsize}
\usepackage{tikz}
\usepackage{longtable}
\usepackage{booktabs}
\usepackage{marvosym} 
\usepackage{framed}
\usepackage{mdwlist}
\usepackage{tabularx}
\usepackage{wrapfig}
\usepackage{array}
\usepackage{mathtools}
\usepackage{lscape}
\usepackage{afterpage}
\usepackage[bookmarks=false,draft]{hyperref}
\usepackage{algorithm}
\usepackage{algpseudocode}
\hypersetup{ colorlinks,
  linkcolor=darkblue,
  filecolor=darkgreen,
  urlcolor=darkred,
  citecolor=darkblue }
\usepackage{varwidth}

%% file: macros.tex
\definecolor{darkblue}{rgb}{0.0, 0.0, 0.55}
\definecolor{cornellred}{rgb}{0.7, 0.11, 0.11}
\definecolor{gray}{rgb}{0.5,0.5,0.5}
\definecolor{purple}{rgb}{0.58,0,0.82}
\definecolor{bsgreen}{RGB}{104, 159, 56}
\definecolor{maroon}{rgb}{0.5,0,0}
\definecolor{darkgreen}{rgb}{0,0.5,0}

\newcommand{\Ignore}[1]{}

\newcommand{\ie}{i.e.}

\newcommand{\eg}{e.g.}
\newcommand{\etal}{et al.}

\newcommand{\CodeIn}[1]{{\small\texttt{#1}}}

\newcommand{\tab}{\hspace{1cm}}

\lstdefinelanguage{XML}
{
  basicstyle=\ttfamily\footnotesize,
  morestring=[b]",
  moredelim=[s][\bfseries\color{maroon}]{<}{\ },
  moredelim=[s][\bfseries\color{maroon}]{</}{>},
  moredelim=[l][\bfseries\color{maroon}]{/>},
  moredelim=[l][\bfseries\color{maroon}]{>},
  morecomment=[s]{<?}{?>},
  morecomment=[s]{<!--}{-->},
  commentstyle=\color{darkgreen},
  stringstyle=\color{blue},
  identifierstyle=\color{red}
}

%% file: intro.tex
\section{Introduction}
\label{sec:intro}

The Internet of Things (IoT) is growing rapidly, with 8 billion
of ``things'' connected in 2017, and 20 billion expected by
2020. Among these, the consumer segment is the largest, with 63\% of installed devices in
2017, in contrast with business-specific domains
~\cite{gartner-device-numbers}. Security has been a key concern, with research focusing on both hardware (smart
devices) and software (IoT frameworks/platforms) to avoid events like
the Mirai Botnet attack, which affected 100,000 IoT devices around the world ~\cite{mirai-attack}.


Mobile applications targeting IoT devices are commonly developed using
IoT frameworks, which provide a set of guiding protocols and standards
that simplify the implementation of IoT applications. They enable
developers to build apps that compute on data emitted by IoT devices
(\eg{}, camera, heart-rate, and temperature sensors). IoT devices
typically rely on cloud services to allow users to monitor and control
the devices remotely from their smartphones, requiring users to
authenticate to the cloud services or, in some cases, to the device and,
under typical permission models used for apps, to grant the app full
access to the Internet. 

This permission model
is too permissive from a security standpoint as it controls
\textit{what} sources and sinks the app can access, but not
\textit{how} information flows between sources and sinks. For example,
a permission which grants the permission to send camera data (source)
to the network (sink) can leak data to arbitrary malicious servers, as the
permission does not state which flows of this kind are allowed. In the
case of sensitive user-interface (UI) data, such as userid/passwords for
cloud services or devices, existing platforms do not even provide a mechanism to tag the UI data as sensitive and thus
the sensitive UI data can be arbitrarily leaked to any authorized sink.

Fernandes \etal{}~\cite{flowfence:16} proposed in prior work
FlowFence, a framework based on taint analysis which forces an app
developer to declare--through flow policies--the intended flows of
sources and sinks allowed for that app. These policies are dynamically
enforced by FlowFence, which checks whether a flow from source to sink
is allowed. While FlowFence enables developers to secure data flows
between sources and sinks, it did not provide (1) ability to declare
User-Interface (UI) fields as sensitive sources and (2) ability to
constrain the network addresses to be used as sinks.

To illustrate the importance of these features, consider, for example,
a sensitive UI field in the app, say a password field, that is used to
authenticate into a legitimate server. Unfortunately, FlowFence
cannot prevent an unwanted flow of this field to a \textit{NETWORK}
sink. This is because (1) the password field could not be declared as
a sensitive source; and (2) even if the password field was declared as
sensitive source, FlowFence would allow the flow anyway, as it cannot
enforce flow policies to \textit{NETWORK} sinks in a fine-grained
manner, \ie{}, a declared flow of \textit{SOURCE->NETWORK} permit a
flow of \textit{SOURCE} to every server on the \textit{NETWORK}, not
just some specific server.


The fundamental problem is that current solutions cannot control
information flows between sources and sinks in a fine-grained
manner. Also, they do not treat UI values as sensitive sources, as to
avoid leaking these values to unwanted sinks (\eg{}, ~\textit{SMS, NETWORK}).

The main contributions of this work are: (1) we enable developers to
tag UI fields holding sensitive data as sensitive sources, enforcing
flow policies upon those sources to sinks; (2) we provide
mechanisms for declaring flow policies that secure network requests to
custom endpoints; and (3) though our work was motivated by (in)security of IoT apps,
our prototype is Android-based and broadly applicable for helping developers
protect sensitive credentials in other Android apps as well.

%% file: motivation.tex
\section{Motivating Example}
\label{sec:motivation}

\begin{figure}[ht]
\centering
\includegraphics[width=.5\textwidth]{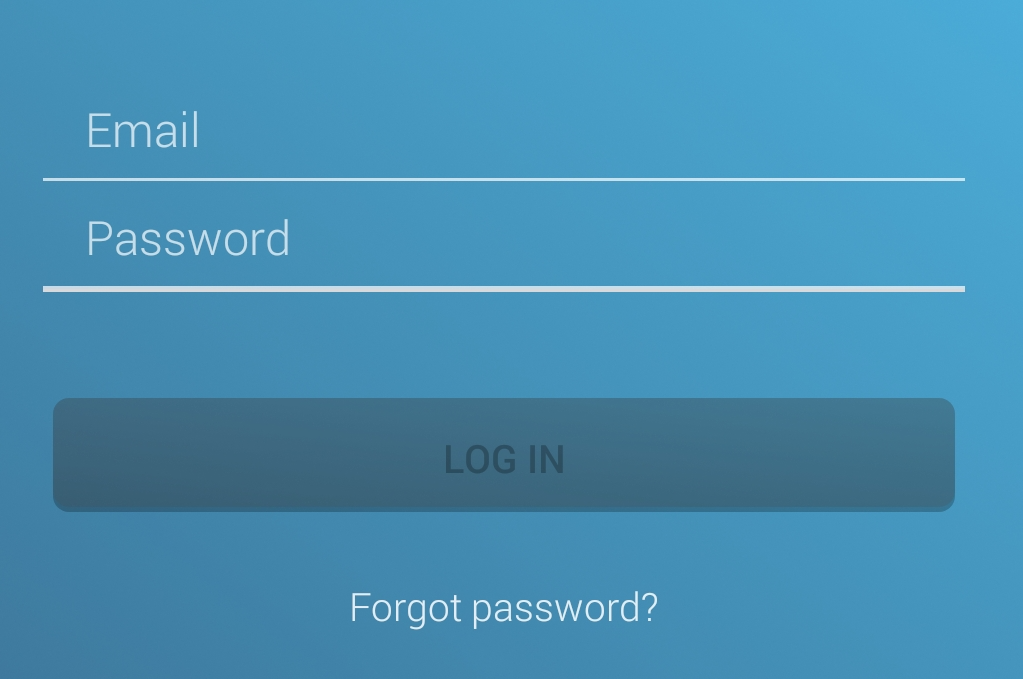}
\captionsetup{justification=centering}
\caption{A commonly seen UI screen in IoT apps. User enter his
  credentials so he can login and access his IoT device.}
\label{fig:login_screen}
\end{figure}

To better understand the importance of securing UI values and network
requests in apps, consider the following scenario. Figure~\ref{fig:login_screen} illustrates a screen that may be displayed when a user wishes
to control an IoT device remotely. After looking at 32 consumer
best-selling IoT brands including Nest, Ring, August Home and TP-Link,
we found that all of them presented a login screen. The user is prompted to enter his credentials and authenticate to the
device manufacturer's cloud server so he can access IoT devices that
were registered on the user's account. These credentials usually include an email and password,
consisting of sensitive data that, if leaked, could compromise the IoT device
being accessed.

On Android, the app needs permission to access \textit{NETWORK}, as
the login process involves validating the user credentials to a
server. This permission is too coarse-grained, as the data could be
sent to any untrusted server if the app was compromised.

Figure \ref{lst:smart-app} shows pseudocode of an IoT app
illustrating this scenario. Line 2 shows a reference to default
Android UI components (\CodeIn{EditText}) used to get the email and
password values they hold. For that, one needs to call the
\CodeIn{getText} method on both components (lines 3 and 4). Then, the
values are used as parameters of another method call that continues
the login process (line 5). The process of getting UI values is pretty
simple, and as such, could be exploited. For example, taking advantage
of the Android UI components, an adversary could inject malicious code
that replicates line 3 and 4, but instead of sending the credentials
to a cloud server, leaks the data to an untrusted server.

\begin{figure}[ht]
  \centering
\noindent\fbox{%
\begin{varwidth}{\dimexpr\linewidth-2\fboxsep-2\fboxrule\relax}
\begin{algorithmic}[1]
\footnotesize
  \State \textbf{application} SmartApp
  \State EditText emailUI, passwordUI;
  \State email = emailUI.getText();
  \State password = passwordUI.getText();
  \State login(email, password);
\end{algorithmic}
\end{varwidth}%
}
\captionsetup{justification=centering}
\caption{Pseudocode of a Smart App using default UI components.}
  \label{lst:smart-app}
\end{figure}

%% file: related.tex
\section{Related Work}

Inherited from the smartphone domain, current IoT frameworks use a
permission-based model to restrict which resources mobile apps can
access. Previous studies demonstrated how this model is inadequate
due to its coarse-grained control of application permissions.
Analyzing the permission system of the Android platform, Feng~\etal{} presented several issues and attacks that took advantage of
the permission-based model, describing over-claim of permissions as
the most severe threat to Android security~\cite{fang:14}. A study
conducted by Felt~\etal{} investigated the requested permissions of 940 apps in the
Android platform, demonstrating that 94\% of the analyzed apps had 4
or more extra permissions that were not used by the app~\cite{felt:11}. 

Previous studies demonstrated how third-party libraries were being
used as attack vectors on mobile apps, taking advantage of the app
broad's access permissions~\cite{sun-gang:14,backes-bugiel-derr:16}. For
instance, Backes \etal{} conducted a study on the top-downloaded
Android apps of Google Play, concluding that 296 of these apps, which had a
cumulative install base of 3.7 billion devices, were using
advertisement third-party libraries with well-known security
vulnerabilities~\cite{backes-bugiel-derr:16}.

The fundamental problem of the permission model resides on information
flows not being controlled once permission to access different sources
of data are granted on the apps. To mitigate this problem, previous
studies used information flow techniques to identify potential
threats. Some of those techniques used static and dynamic analysis to
tackle the problem~\cite{droidsafe:15,taintdroid:14,flowdroid:14,
  phosphor:14}. For instance, Gordon~\etal{} presented DroidSafe, a
static analyzer that used taint tracking to identify malicious flows
between sources and sinks upon installation of the app. Albeit
effective, techniques like DroidSafe have problems dealing with
dynamic code injection and implicit flows~\cite{droidsafe:15}. Other
techniques proposed were based on flow policies to control information
flow on the program~\cite{ fine-grained-policy:10, flowfence:16}. 

Building on the idea of using flow policies to control data flows
within the app, Fernandes~\etal{} presented FlowFence, a framework that
enables the development of secure apps under the Android
platform~\cite{flowfence:16}. Its main concept relies on developers
making the app's use patterns explicit through flow policies. These
policies are then enforced by FlowFence to control flows of sensitive
sources to sinks within the app~\cite{flowfence:16}. Albeit proving to
be an effective solution, FlowFence did not cover UI as sensitive
sources. Also, it did not provide mechanisms to control network
requests in a fine-grained manner. In the following, we briefly
discuss FlowFence's design in Section~\ref{sec:background-flowfence}
and how we extend FlowFence to cover such scenarios in
Section~\ref{sec:approach}.

%% file: background-flowfence.tex
\section{Background of FlowFence}
\label{sec:background-flowfence}

FlowFence is a framework built for mobile development that enables
definition and enforcement of information flow policies under the
Android platform. Designed with the IoT architecture in mind,
Fernandes~\etal{} introduced a new information flow model referred
as~\textit{Opacified Computation}. This model enables developers to
tag sensitive data with taint labels within the app structure,
declaring information flow policies bound to these labels. The
rationale is that, once sensitive data is tagged, any computation
using this data must run in a sandbox, which are separated processes
managed by FlowFence. On these sandboxes, taints are automatically
tracked following the defined flow policies. We briefly describe the
architecture of FlowFence and how its main components are used for
enforcement of information flow policies below.

\subsection{FlowFence Architecture}
\label{subsec:flowfence-arch}

FlowFence consists of two major components: (1) Developer-written
functions (Quarantine Modules) that operate on sensitive data inside
FlowFence-created~\textit{sandboxes} and (2) A Service (\textit{Trusted Service}) that manages these
\textit{sandboxes} and mediate data flows between sources and sinks, also
enforcing statically defined policies and providing APIs for accessing
sensitive sinks, \eg{}, \textit{NETWORK, SMS}. 

\textbf{Quarantine Modules.}
Quarantine Modules (QM) consist of developer-written
functions that operate on sensitive data and execute only in
FlowFence-created \textit{sandboxes}, \ie{}, processes created and
managed by the \textit{Trusted Service}. These QM functions take
serializable data as parameters and return \textit{opaque handles},
which are references to sensitive data that can only be declassified by other
QMs or via trusted service.

\textbf{Trusted Service/API.}
A service responsible for: (1) creation and management of
\textit{sandboxes}; (2) control of all data flows between QMs tagged
by taint labels, enforcing policy rules linked to these labels; and
(3) providing APIs to sensitive sinks that can only be used within
QMs.


\textbf{Flow Policies.}
Flow policies are declared in the app manifest. For every sensitive
data, a taint label is defined in the form \textit{(appID, name)},
where \textit{appID} is the unique identifier of the app (represented by
package name in the FlowFence implementation) and the \textit{name} being
the taint label itself. Within the declared label, a sensitive
source of data is defined with its intended flows in the form
\textit{TaintLabel->Sink}. For example, one could define a taint label
\textit{(com.package.camera\_app, TaintCamera)} and declare its intended
flows, \eg{}, \textit{TaintCamera->Network}. 


%% file: approach.tex
\section{Our Approach}
\label{sec:approach}

One of the key challenges of developing secure apps is that there is
no official security guidelines to obtain UI data from the
user. Developers often need to send this data through the network, so
securing data flows between sensitive UI sources and network sinks is
crucial. Broadly speaking, our approach is developer-driven in the
sense that it enables developers to secure data flows involving
sensitive UI fields and network requests against dynamic code execution, \eg{},
malicious third-party libraries. To describe our approach, we
use the scenario presented in Section~\ref{sec:motivation} throughout the
rest of this section.

\subsection{Sensitive UI.}
\label{subsubsec:ui-sensitive-source}
To secure sensitive UI data, we extend FlowFence so that UI components
can be defined as sensitive sources. For that, we developed new UI components (SensitiveUI) that extend the Android SDK
ones such that default behavior remains intact, but accessing the
value of the components can only be done in a secure manner. Consider an Android UI component
like \CodeIn{EditText}, for example. To \textit{get} or
\textit{set} its value programmatically, one needs to call the
\CodeIn{getText/setText} method in a reference object to that
component. As the created SensitiveUI components extend the
Android-based ones, they include both methods. However,
accessing these methods is considered sensitive code and 
needs to be executed within a FlowFence QM. For instance, trying to
access the \CodeIn{getText} without using FlowFence would return an
empty value.

Inspired by FlowFence, we use a Key-Value mechanism to store UI values. Each
SensitiveUI component has its value associated with an ID and taint
label in the form (\CodeIn{<id, sensitive\_value, taint\_label>}), with
\CodeIn{taint\_label} being declared in a developer-written QM and serving
the purpose of tagging the sensitive data to enforce information flow
policies.

Figure~\ref{lst:smart-app-secure} shows pseudo-code of the example
app, only this time using the SensitiveUI component instead of the
default Android one. Because we build on FlowFence, we need to split
sensitive code into QMs as described in Section
~\ref{sec:background-flowfence}. Lines 4-6 show a QM that calls the
Trusted API to obtain UI values. FlowFence's infrastructure ensures
that whenever a QM is called, the return value is converted to an
opaque handle. The value that this opaque handle holds can only be
accessed by declassifying the handle within a FlowFence-created
sandbox, thus running in a secure environment. Line 8 shows another QM
responsible for the login process, receiving as parameter opaque
handles holding a reference to email and password values. Lines 11 and
12 show how the UI values are recovered, with the QM being
called. Finally, line 13 shows both email and password opaque handles
being used as parameters to the QM responsible for the login. Notice that
calling the \CodeIn{getText} method upon any SensitiveUI reference is
pointless (line 14), as the value can only be accessed within a QM. By
using this component, the developer can specify which flows are
permitted using the sensitive UI values as sources with flow
policies. FlowFence would then proceed to enforce these policies while
blocking all undeclared flows that use sensitive UI values.

\begin{figure}[ht]
\centering
\noindent\fbox{%
\begin{varwidth}{\dimexpr\linewidth-2\fboxsep-2\fboxrule\relax}
\begin{algorithmic}[1]
  \footnotesize
  \State \textbf{application} SmartApp
  \State SensitiveUI emailUI, passwordUI;
  \State
  \State \textbf{String} QM\_getUIValue(id):
  \State \tab value = TrustedAPI.sensitiveUI.getText(id);
  \State \tab \textbf{return} value;
  \State
  \State \textbf{void} QM\_login(email, password):
  \State \tab \textit{// Continue the log in process}
  \State
  \State email = \textbf{QM.call} (QM\_getUIValue, emailUI.id);
  \State password = \textbf{QM.call} (QM\_getUIValue, passwordUI.id);
  \State \textbf{QM.call} (QM\_login, email, password);
  \State password = passwordUI.getText(); \textit{// This call returns an empty value}
\end{algorithmic}
\end{varwidth}%
}
\captionsetup{justification=centering}
\caption{Pseudo-code of a Smart App using secure UI components in a login screen.}
\label{lst:smart-app-secure}
\end{figure}

\subsection{Secure Network Requests.}
\label{subsubsec:secure-network}
                
To make secure network requests possible using FlowFence
infrastructure, we start by extending the Trusted API so it
can execute network requests to custom URLs. For that, we extend
FlowFence's flow policy language to allow
filtering of custom URLs through the definition of flow policies in a
fine-grained manner, which we discuss next.

With FlowFence original implementation, the developer
could only specify a flow of \textit{SOURCE} to \textit{NETWORK},
therefore permitting flows to any URL. Now, the developer can also specify which URL it
wants the source data to sink to, declaring the rule in the form
(\textit{TaintLabel -> NETWORK, URL}). FlowFence policy checker would
then compute the rule, also checking the URL of the request,
either granting or denying the flow.


Figure~\ref{lst:smart-app-network-secure} shows pseudocode of the
example app described earlier. Line 3 shows the definition of a policy
that specifies a flow between UI sources and network sink, but only to
a specific cloud server’s URL. Line 11 shows how the Trusted API is
used to make a network request responsible for logging in. After
obtaining opaque handles referencing the user’s credentials (line 13
and 14), the QM responsible for login is called passing the
credentials, server’s URL, and taint label as parameters. With the
Trusted API, the URL and taint labels are not considered when checking
which network flows are permitted on the app. Line 16 shows how an
unauthorized request to an untrusted server would be denied, as
FlowFence policy checker would validate the requested URL to the
declared flow policy.

\begin{figure}[ht]
\centering
\noindent\fbox{%
\begin{varwidth}{\dimexpr\linewidth-2\fboxsep-2\fboxrule\relax}
\begin{algorithmic}[1]
  \footnotesize
  \State \textbf{application} SmartApp
  \State taint\_label Taint\_UI;
  \State allow \{ Taint\_UI -> NETWORK, http://appcloudserver.com \}
  \State SensitiveUI emailUI, passwordUI;
  \State
  \State \textbf{String} QM\_getUIValue(id):
  \State \tab value = TrustedAPI.sensitiveUI.getText(id);
  \State \tab \textbf{return} value;
  \State
  \State \textbf{void} QM\_login(email, password, url):
  \State \tab TrustedAPI.network.post(email, password, url);
  \State
  \State email = \textbf{QM.call} (QM\_getUIValue, emailUI.id);
  \State password = \textbf{QM.call} (QM\_getUIValue, passwordUI.id);
  \State \textbf{QM.call} (QM\_login, email, password,
  http://appcloudserver.com, Taint\_UI);
  \State \textbf{QM.call} (QM\_login, email, password,
  http://untrustedserver.com, Taint\_UI); \textit{// This request
    would be denied, as there is no policy specifying a flow between
    UI source and this URL as sink.}
\end{algorithmic}
\end{varwidth}%
}
\captionsetup{justification=centering}
\caption{Pseudo-code of a Smart App using secure UI components to send
  data through the network.}
\label{lst:smart-app-network-secure}
\end{figure}

Let's reconsider the scenario in Section 2 and show how the above
mechanisms help reduce the attack surface available to untrusted
third-party libraries.  Malicious code that could be executed either
by third-party libraries or through dynamic code injection outside a
QM would not gain read access to sensitive UI data in the
FlowFence-protected sandbox due to standard FlowFence mechanisms. Even
if injected code manages to execute inside a QM, it would not be able
to leak intercepted credentials to arbitrary servers due to fine-grain
network policy. A limitation of our work that remains is that we do
not prevent phishing attacks that use a fake UI screen generated from
outside a QM by injected code. We plan to address the limitation in
our future work. A possible approach is to extend the mechanisms for
protecting against UI deception described
in~\cite{fernandes-et-al:16,bianchi-et-al:15} to assist the user in
distinguishing the sensitive UI fields that are generated from within
a QM and those that are not.

%% file: conclusion.tex
\section{Conclusions}
\label{sec:conclusion}

Controlling how data flows within IoT apps in a fine-grained manner is
crucial to avoid data leakage. In this work, we presented our work in
progress on a solution for the limitations of the permission-based model
which are often used in current IoT and smartphone
frameworks. We addressed
how developers can secure sensitive sources of UI and control network flows in a
fine-grained manner. As future work, we envision the evaluation of our
approach with developers, quantifying both effort to port IoT apps as
well as performance impact. We also plan to develop further ideas for
controlling information flows within mobile apps.

%% file: acknowledgements.tex
\section{Acknowledgements}
\label{sec:ack}

This reserach is supported by NSF under grants No. 1740897 and
1740916, and by RNP under grant No. 002951. The authors thank the anonymous reviewers and also Luís Melo and Harvey Lu
as well as professors Darko Marinov and Marcelo d’Amorim for their
valuable feedback.